\documentclass[ floatfix, amsmath, amssymb, aps, prd, twocolumn, lengthcheck, sort&compress, showpacs,  nofootinbib, superscriptaddress ]{revtex4-1}

\usepackage{graphicx}
\usepackage{dsfont}
\usepackage{dcolumn}
\usepackage{units}
\usepackage{wasysym}
\usepackage{multirow}
\usepackage{slashed}
\usepackage[usenames]{color}
\usepackage[colorlinks=true,linkcolor=blue,citecolor=blue,urlcolor=blue]{hyperref}
\usepackage{tabularx}
\usepackage{booktabs}

\usepackage[usenames]{color}
\usepackage{colortbl}

\definecolor{violet}{RGB}{111,0,255}
\definecolor{webgreen}{rgb}{0,0.75,0}
\definecolor{webred}{rgb}{0.75,0,0}
\definecolor{webblue}{rgb}{0,0,0.75}
\definecolor{darkblue}{rgb}{0,0,0.6}
\definecolor{darkgreen}{rgb}{0,0.5,0.5}
\definecolor{darkpurple}{rgb}{0.5,0,0.5}
\definecolor{darkorange}{rgb}{1,0.5,0}
\definecolor{darkgrey}{rgb}{0.4,0.4,0.4}
\definecolor{lgray}{rgb}{0.95,0.95,0.95}
\definecolor{lgreen}{rgb}{0.95,1.00,0.90}
\definecolor{lred}{rgb}{1.00,0.90,0.80}
\definecolor{lblue}{rgb}{0.2,0.35,1.00}
\definecolor{shadecolor}{rgb}{1.00,0.92,0.82}

\usepackage{bm}

\newcommand{\conjg}[1]{\ensuremath{\hspace{1pt}\overline{\hspace{-1pt}#1\hspace{-1pt}}}\hspace{1pt}}

\def\Slash#1{\setbox0=\hbox{$#1$} 
\dimen0=\wd0 
\setbox1=\hbox{/} \dimen1=\wd1 
\ifdim\dimen0>\dimen1 
\rlap{\hbox to \dimen0{\hfil/\hfil}} 
#1 
\else 
\rlap{\hbox to \dimen1{\hfil$#1$\hfil}} 
/ 
\fi}

\newcommand{\beq}{\begin{equation}}
\newcommand{\eeq}{\end{equation}}
\newcommand{\beqa}{\begin{align}}
\newcommand{\eeqa}{\end{align}}

\begin{document}

\title{On light baryons and their excitations}

\author{Gernot Eichmann}
\email{gernot.eichmann@physik.uni-giessen.de}
\affiliation{Institut f\"ur Theoretische Physik, Justus-Liebig--Universit\"at Giessen, 35392 Giessen, Germany.}

\author{Christian S. Fischer}
\email{christian.fischer@physik.uni-giessen.de}
\affiliation{Institut f\"ur Theoretische Physik, Justus-Liebig--Universit\"at Giessen, 35392 Giessen, Germany.}
\affiliation{HIC for FAIR Giessen, 35392 Giessen, Germany}

\author{H\`elios Sanchis-Alepuz}
\email{helios.sanchis-alepuz@uni-graz.at}
\affiliation{Institute of Physics, NAWI Graz, University of Graz, Universit\"atsplatz 5, 8010 Graz, Austria}

\begin{abstract}
We study ground states and excitations of light octet and decuplet baryons within the
framework of Dyson-Schwinger and Faddeev equations. We improve upon similar approaches by
 explicitly taking into account the momentum-dependent dynamics of the quark-gluon interaction
that leads to dynamical chiral symmetry breaking. We perform calculations in both
the three-body Faddeev framework and the quark-diquark approximation in order to assess the
impact of the latter on the spectrum. Our results indicate that both approaches agree well
with each other. The resulting spectra furthermore agree one-to-one with experiment, provided
well-known deficiencies of the rainbow-ladder approximation are compensated for.
We also discuss the mass evolution of the Roper and the excited $\Delta$ with varying pion mass
and analyse the internal structure in terms of their partial wave decompositions.
\end{abstract}


\maketitle

\section{Introduction}
Understanding the baryon excitation spectrum of QCD
is one of the key elements in unravelling the structure of the strong interaction.
In the past years, significant experimental progress has been made by the analysis
of data from photo- and electroproduction experiments at JLAB, ELSA and MAMI
\cite{Aznauryan:2009mx,Aznauryan:2012ba,Sarantsev:2007aa}. As a result,
a number of new radial and orbital excitations of the ground state
octet baryons have been added to the PDG \cite{Agashe:2014kda}.

Despite this progress, there are still longstanding issues with the
baryon spectrum that are not well understood. One of these is the prediction of
many excited states by the quark model which, however, have not been observed yet.
This `missing resonances' problem has been debated at length in the literature
but remains an open issue.
One of the proposed solutions has been a quark-diquark picture
of baryons, with a strongly bound and therefore hard to excite diquark that
prevents the appearance of many states present in the constituent three-quark
model, see~\cite{Anselmino:1992vg,Klempt:2009pi} for reviews. In practice, it may be
hard to reconcile such strongly bound diquarks with the underlying QCD forces and
it is a non-trivial question whether baryons made of loosely correlated diquarks
with non-trivial internal structure can be distinguished in their spectrum from
genuine three-body states. This is one of the topics of this work.

Another longstanding issue is the level ordering
between the first radially excited state in the $I(J^P) = \frac{1}{2}(\frac{1}{2}^+)$
sector (the Roper resonance \cite{Roper:1964zza}) and the ground state in the orbitally excited channel
$I(J^P) = \frac{1}{2}(\frac{1}{2}^-)$: whereas quark model calculations typically
favour the `natural' ordering of a lower $\frac{1}{2}^-$ state
\cite{Isgur:1978wd,Glozman:1995fu,Loring:2001kx,Loring:2001ky},
the measured mass of the Roper is much lower than the quark model prediction and
the level ordering reversed. Thus it was conjectured that the inner structure
of the Roper may be more complicated than that of a `simple' radial excitation. Further
indications in this direction may be inferred from its large decay width
and the large branching fractions in the $\pi N$ and $\sigma N$ decay channels.

An interesting possibility connected with the latter observation has been discussed
in \cite{Suzuki:2009nj}: a radial excitation of the nucleon with an initial (or `bare') mass
much larger than the experimental one may receive large corrections from coupled
channel effects in the $\pi N$, $\pi \pi N$ and $\eta N$ channels. The resulting
mass of the Roper resonance observed in the data is then substantially lowered and
may be pushed below the one of the negative parity ground state. In this
picture the internal structure and the properties of the resulting dressed state may
be very different than those of the initial bare state. In fact, the reaction
dynamics may  even be strong enough to generate the Roper purely dynamically without
a bare seed, as demonstrated in \cite{Doring:2009yv,Bruns:2010sv,Liu:2016uzk}.

Additional insight into the nature of the Roper may be gained from lattice QCD.
However, excited states in general pose a challenge for the lattice as the extraction
of their masses from Euclidean correlators is an intricate statistical problem.
The computational cost involved in unquenched simulations of excited states
often necessitates the use of unphysically heavy light quark masses. In addition, the
spectrum is complicated by the appearance of discrete multi-particle scattering states
generated by the finite volume on the lattice. As a consequence, contemporary lattice
data from several groups on the mass evolution of the first radial excitation of the
nucleon seem to differ both quantitatively and qualitatively \cite{Mahbub:2010rm,Edwards:2011jj,Alexandrou:2013fsu,Roberts:2013ipa,Alexandrou:2014sha,Liu:2014jua,Alexandrou:2015hxa}.

            \begin{figure*}[t]
            \centerline{%
            \includegraphics[width=0.70\textwidth]{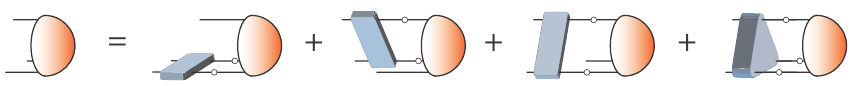}}
            \caption{Three-quark Faddeev equation.}
            \label{fig:faddeev}
            \end{figure*}

In the framework of Dyson-Schwinger, Bethe-Salpeter and Faddeev equations
(see~\cite{Roberts:1994dr,Alkofer:2000wg,Fischer:2006ub,Bashir:2012fs,review} for reviews)
the properties of the Roper resonance have been analysed so far on two levels
of sophistication~\cite{Roberts:2011cf,Roberts:2011ym,Chen:2012qr,Segovia:2015hra}.
Both of these approaches
convert the three-body system into a
two-body quark-diquark picture assuming strong quark-quark correlations inside
baryons. In addition,
one of them~\cite{Roberts:2011cf,Roberts:2011ym,Chen:2012qr}
employs an NJL-type, momentum-independent vector-vector interaction between the quarks
which leads to momentum-independent wave functions for the diquark constituents and the
resulting baryons, thereby neglecting parts of the underlying QCD dynamics. In a recent
second study~\cite{Segovia:2015hra} momentum-dependent model ans\"atze for the quark
propagator and the diquark wave functions were employed to study their impact
on the properties of the nucleon's excitations. While both approaches agree in
their general conclusions, they differ considerably in their description of the
internal properties of the Roper resonance.

In this work we improve upon the situation in two respects. First, we use
a well-established momentum-dependent effective quark-gluon interaction as a starting
point and determine all propagators and wave functions self-consistently from their
Dyson-Schwinger and Bethe-Salpeter equations (DSEs and BSEs). This procedure serves to eliminate unwanted
freedom in modelling and takes care of the preservation of chiral symmetry via the axial
Ward-Takahashi identity. Second, we do not rely on the quark-diquark approximation.
Instead, we provide first solutions for the excited state spectrum of the
three-body Faddeev equation. In parallel, we also solve the bound-state equation
for baryons in the quark-diquark system using the same underlying interaction.
We are therefore in a position to systematically compare the results in both approaches
and assess their qualitative and quantitative differences. We apply this formalism
to nucleon and $\Delta$ baryons with quantum numbers $J^P=1/2^\pm$
and $3/2^\pm$ and discuss implications for the interpretation of the experimental
spectrum.

This work is organized as follows. In the next section we discuss the details of
the three-body Faddeev approach to baryons and
the quark-diquark BSE and we specify the quark-gluon interaction used in this work. We present
and discuss our results in Sec.~\ref{sec:results} and conclude in Sec.~\ref{sec:conclusions}.

\section{Three-body Faddeev equations vs. quark-diquark approximation}\label{sec:form}

In functional frameworks the masses and wave functions of baryons are extracted from
their gauge-invariant poles in the (gauge-dependent) quark six-point Green function. There is an intimate
relation to the corresponding procedure in lattice QCD, see Ref.~\cite{review} for detailed explanations.
As a result, one arrives at the covariant three-body Faddeev equation in Fig.~\ref{fig:faddeev}
which is an exact equation in QCD.
It determines the baryon's three-quark Faddeev amplitude from the irreducible
two- and three-body interactions between the dressed valence quarks.
It is also much more complicated than the analogous two-body equations for mesons,
partially due to the structure of the baryon amplitude
which depends on three independent momenta and many more tensor structures than the meson case.
Sophisticated methods based on permutation-group symmetries have been developed
to deal with the complexity of this equation, see Refs.~\cite{Eichmann:2011vu,Eichmann:2015nra} for
state-of-the-art solution techniques.

The structure of the baryon Faddeev amplitudes is discussed in Ref.~\cite{Eichmann:2009qa,Eichmann:2011vu,SanchisAlepuz:2011jn,review}
and shall not be explicitly repeated here for brevity. For later use we just state that the 64 different tensor
structures representing a $J=1/2$ baryon can be grouped into eight $s$-wave components,
36 $p$ waves and 20 $d$ waves. Analogously, the 128 tensor structures of a $J=3/2$ Faddeev
amplitude comprise four $s$ waves, 36 $p$ waves, 60 $d$ waves and 28 $f$ waves. The
multiplicity of these components is certainly not in one-to-one relation with their relative importance
in the baryon's amplitude; we come back to this issue in the results section below.

            \begin{figure*}[t]
            \centerline{%
\includegraphics[width=0.95\textwidth]{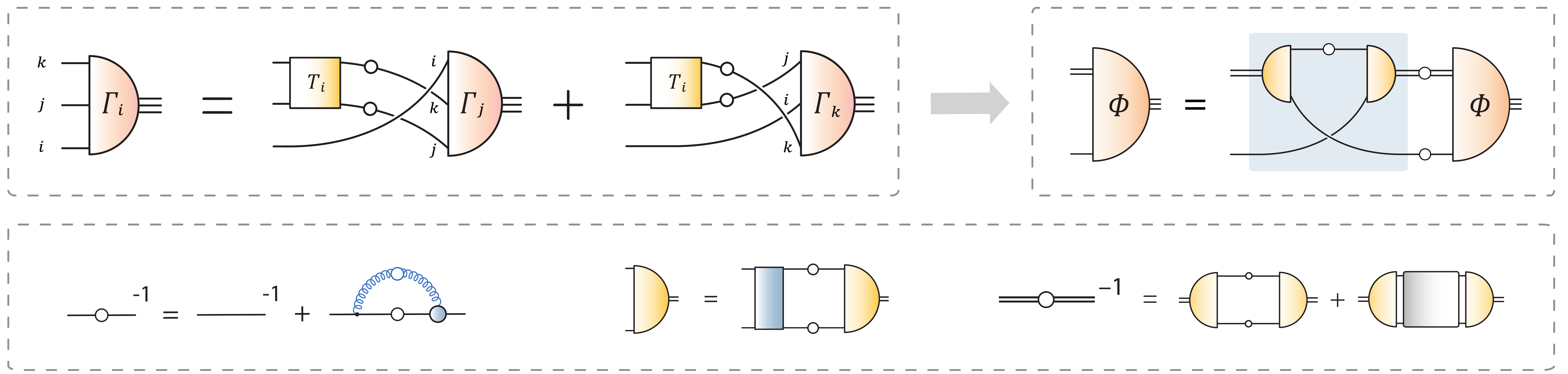}}
            \caption{Simplification of the Faddeev equation in Eq.~\eqref{fe-rewritten} (\textit{top left}) to the quark-diquark Bethe-Salpeter equation~\eqref{qdq-bse} (\textit{top right}).
                     The bottom panel shows the ingredients that enter in the equation and are calculated beforehand: the quark propagator, diquark Bethe-Salpeter amplitudes and diquark propagators.}
            \label{fig:quark-diquark}
            \end{figure*}

The Faddeev equation in Fig.~\ref{fig:faddeev} contains irreducible three-body forces in the last
diagram on the r.h.s. whose influence on the spectrum has not yet been fully
explored. However, from a diagrammatic viewpoint it seems plausible that they only play a minor role.
The leading diagram in a skeleton expansion is one with a dressed three-gluon-vertex
with gluon propagators attached to each of the three quarks. It has been shown,
however, that this contribution vanishes trivially due to the colour algebra, with the leading non-trivial terms
identified and explored in~\cite{Sanchis-Alepuz:threebodyinprep}. Therefore it seems not unreasonable
to neglect irreducible three-body forces altogether and evaluate the three-body problem with two-body interactions
only. This is the strategy followed so far in Refs.~\cite{Eichmann:2009qa,Eichmann:2011vu,Eichmann:2011pv,SanchisAlepuz:2011jn,Sanchis-Alepuz:2013iia,Sanchis-Alepuz:2014wea,
Sanchis-Alepuz:2014sca,Sanchis-Alepuz:2015fcg,Sanchis-Alepuz:2015qra} and we will also adopt it in
this work. As will become apparent in Sec.~\ref{sec:results}, our results will justify this approximation
a posteriori.
The Faddeev equation then takes the form
           \begin{equation}\label{faddeev-eq}
             \mathbf{\Gamma} = \sum_{i=1}^3 \mathbf\Gamma_i = \sum_{i=1}^3 \mathbf{K}_i\,\mathbf{G}_0\,\mathbf{\Gamma}\,,
           \end{equation}
           where the Faddeev components $\mathbf\Gamma_i$ correspond to the three individual diagrams in Fig.~\ref{fig:faddeev}.
           The $\mathbf{K}_i$ are the two-body kernels and $\mathbf{G}_0$ refers to the product of two quark propagators.
           Eq.~\eqref{faddeev-eq} constitutes the first of the two frameworks that we employ below to calculate the baryon spectrum.

           The second is the quark-diquark approach, which is motivated by the assumed smallness of irreducible three-body contributions.
           We can eliminate the two-body kernels in Eq.~\eqref{faddeev-eq} in favor of the two-body T-matrices $\mathbf{T}_i$, which are related to each other via Dyson's equation:
           \begin{equation}\label{dyson-eq-2}
             \mathbf{T}_i = (\mathds{1}+\mathbf{T}_i\,\mathbf{G}_0)\,\mathbf{K}_i\,.
           \end{equation}
           Applying this to the Faddeev equation gives
           \begin{equation}\label{fe-rewritten}
           \begin{split}
             \mathbf{T}_i\,\mathbf{G}_0\,\mathbf{\Gamma} &= (\mathds{1}+\mathbf{T}_i\,\mathbf{G}_0)\,\mathbf{\Gamma}_i \quad \Rightarrow \\
             \mathbf\Gamma_i &= \mathbf{T}_i\,\mathbf{G}_0\,(\mathbf\Gamma-\mathbf\Gamma_i) = \mathbf{T}_i\,\mathbf{G}_0\,(\mathbf\Gamma_j+\mathbf\Gamma_k)
           \end{split}
           \end{equation}
           with $\left\{i,j,k\right\}$ an even permutation of $\left\{1,2,3\right\}$.
           The resulting equation is shown in the upper left panel of Fig.~\ref{fig:quark-diquark};
           so far no further approximation has been made.
           However, its structure motivates to expand the quark-quark scattering matrix that appears therein in terms of separable diquark correlations.
           The sum of diquarks is then dominated by those with smallest
           mass scales, namely, scalar and axialvector diquarks in the positive parity sector as well as pseudoscalar and vector diquarks with negative parity.
           In our calculations below we take all of these into account.
The quark-quark scattering matrix then reads :
\begin{equation}
\begin{split}
           &\big[\mathbf{T}(q,q',p_d)\big]_{\alpha\gamma;\beta\delta} \simeq \sum  \big[\mathbf\Gamma^{\mu\ldots}_\text{D}(q,p_d)\big]_{\alpha\beta}  \\
           & \qquad\qquad\,\times D^{\mu\ldots\nu\ldots}(p_d^2)\,  \big[\conjg{\mathbf\Gamma}^{\nu\ldots}_\text{D}(q',p_d)\big]_{\delta\gamma}\;, \label{diquark-approximation} \\
           &\big[\mathbf\Gamma_i(p,q,P)\big]_{\alpha\beta\gamma\sigma} \simeq \sum \big[\mathbf\Gamma^{\mu\ldots}_\text{D}(q,p_d)\big]_{\alpha\beta} \\
           & \qquad\qquad\,\times D^{\mu\ldots\nu\ldots}(p_d^2) \, \big[\Phi^{\nu\ldots}(p,P)\big]_{\gamma\sigma}\;.
\end{split}
\end{equation}
Here, $\mathbf\Gamma^{\mu\ldots}_\text{D}$ is the diquark Bethe-Salpeter amplitude and
$\overline{\mathbf\Gamma}^{\mu\ldots}_\text{D}$ its charge conjugate;
the diquark propagator is $D^{\mu\ldots\nu\ldots}$; $p_d$ is the diquark momentum and $q$,
$q'$ are the relative quark momenta in the diquark amplitudes. In the second line the same assumption
was made for the Faddeev components, thus introducing the \textit{quark-diquark} Bethe-Salpeter amplitude $\Phi^{\nu\ldots}(p,P)$.

We therefore arrive at a coupled system of quark-diquark bound-state equations~\cite{Cahill:1988dx,Hellstern:1997pg,Oettel:1998bk},
which are illustrated in the upper right panel of Fig.~\ref{fig:quark-diquark}:
           \begin{equation}\label{qdq-bse}
           \begin{split}
              &\big[\Phi^{\mu\ldots}(p,P)\big]_{\alpha\sigma} = \int \!\!\frac{d^4k}{(2\pi)^4}\,\big[\mathbf K_\text{Q-DQ}^{\mu\ldots \nu\ldots}\big]_{\alpha\beta} \\
              &\qquad \times \big[S(k_q)\big]_{\beta\gamma}\,D^{\nu\ldots\rho\ldots}(k_d)\,\big[\Phi^{\rho\ldots}(k,P)]_{\gamma\sigma}\,,
           \end{split}
           \end{equation}
           and the quark-diquark kernel is given by
           \begin{equation}
                \mathbf K_\text{Q-DQ}^{\mu\ldots \nu\ldots} =  \mathbf\Gamma_\text{D}^{\nu\ldots}(k_r,k_d)\,S^T(q)\,\conjg{\mathbf\Gamma}_\text{D}^{\mu\ldots}(p_r,p_d)\,.
           \end{equation}
Here, $P$ is the baryon's total momentum, $p$ is the quark-diquark relative momentum and the remaining momenta
can be inferred from the figure (see Sec.~5.2 in~\cite{Eichmann:2009zx} for details). The dressed quark propagator is denoted by $S(q)$ and `T' is a matrix transpose.

In this picture the baryon is bound by quark exchange between the quark and the diquark \cite{Cahill:1988dx,Cahill:1988zi,Reinhardt:1989rw}.
This, however, does not mean that gluons have been eliminated from the problem: they still appear explicitly in the DSE for the quark propagator,
the BSEs for the diquark Bethe-Salpeter amplitudes $\mathbf\Gamma_\text{D}$, and in the equations for the diquark propagators which are all displayed in the bottom panel of
Fig.~\ref{fig:quark-diquark}. The diquark BSEs read
\begin{equation}\label{bse-diquark}
           \begin{split}
            &\left[\mathbf\Gamma_\text{D}^{\mu_1\ldots\mu_J}(p,P)\right]_{\alpha\beta}
            = \int\!\!\frac{d^4q}{\left(2\pi\right)^4} \left[\mathbf{K}(p,q,P)\right]_{\alpha\gamma;\beta\delta} \\
            & \qquad \times\left[ S(q_+)\,\mathbf\Gamma_\text{D}^{\mu_1\ldots\mu_J}(q,P)\,S^T(-q_-)\right]_{\gamma\delta}
            \end{split}
\end{equation}
and contain the same two-body interaction kernel $\mathbf{K}$ as the three-body Faddeev equation.

Although diquarks are not observable because they carry colour,
the rainbow-ladder truncation does generate diquark poles in the two-quark scattering matrix which justifies the approximation~\eqref{diquark-approximation}
and allows one to compute diquark properties in close analogy to those of mesons from their BSEs~\eqref{bse-diquark}.
In a simpler model it has been shown that the addition of crossed ladder exchange removes the diquark poles from the spectrum~\cite{Bender:1996bb}, but
it was recently argued that an effective resummation of such diagrams can also bring them back again~\cite{Jinno:2015sea}.
In any case, diquark correlations may well persist in one form or another simply due to the colour attraction:
it is conceivable that the $qq$ scattering matrix has some complicated singularity structure that allows one to identify diquark mass scales,
and in that sense Eq.~\eqref{diquark-approximation} will remain a reasonable ansatz.

The quark-diquark equation is a considerable simplification, both in terms of kinematic variables and tensor structures.
In turn, much of the complexity is now distributed among the underlying equations which we solve beforehand as described in Refs.~\cite{Eichmann:2007nn,Eichmann:2008ef,Eichmann:2009zx,Nicmorus:2008vb}:
the (scalar, axialvector, pseudoscalar, vector) diquark amplitudes for complex relative momenta including the full set of tensor structures,
and the respective diquark propagators for complex total momenta.
In any case, the rather mild assumptions required to derive the quark-diquark BSE suggest that it may still capture the essential dynamics of the three-body
system, which will be one of the issues that we explore in this work.

The common underlying dynamics of the three-quark and quark-diquark equations is encoded in the two-body scattering kernel $\mathbf{K}$ and, related, the quark-gluon
interaction. The latter also appears in the quark DSE which determines the fully dressed quark propagator $S(p)$.
The quark self-energy is related to the two-body scattering kernel $\mathbf{K}$ via the axialvector Ward-Takahashi identity
which ensures chiral symmetry and, in combination with its correct  dynamical breaking pattern,
the Goldstone-boson nature of the pseudoscalar meson octet. A frequently
used approximation that satisfies this identity is the rainbow-ladder truncation which we also use in this work. Its basic idea
is to approximate all effects of the quark-gluon vertex by its primal tensor structure $\gamma_\mu$ dressed with a function that
depends on the gluon momentum only. This function is then combined with the dressing function of the gluon propagator into a so-called
effective coupling. In the large momentum regime this quantity is well-known from resummed perturbation theory whereas in the infrared
it is modelled. A detailed discussion of rainbow-ladder and other approximation strategies for the DSE/BSE system can be found in Ref.~\cite{review}.

One of the more frequently used effective interactions is that of Maris and Tandy~\cite{Maris:1999nt}:
\begin{equation}
\begin{split}\label{couplingMT}
\alpha(k^2) &= \pi \eta^7  x^2
e^{-\eta^2 x} + \alpha_\text{UV}(k^2) \,,\\
    \alpha_\text{UV}(k^2) &= \frac{2\pi\gamma_m \big(1-e^{-k^2/\Lambda_t^2}\big)}{\ln \, \left[e^2-1+\big(1+k^2/\Lambda^2_{\mathrm{QCD}}\big)^2\right]}
\end{split}
\end{equation}
with $x=k^2/\Lambda^2$.
The UV term with parameters $\Lambda_t=1$~GeV, $\Lambda_\mathrm{QCD}=0.234\,{\rm GeV}$, and $\gamma_m=12/25$ for four active quark flavours
ensures the correct perturbative running but is otherwise not essential;
one could neglect it without causing serious damage in the spectrum of the light-quark sector~\cite{Alkofer:2002bp}.
The nonperturbative physics is encoded in the first term, which is characterised by two parameters\footnote{The
relationship with the parameters $\{ \omega, D \}$ used in Ref~\cite{Maris:1999nt} is $\omega=\Lambda/\eta$ and $D=\eta \,\Lambda^2$.}:
an infrared scale $\Lambda$ and a dimensionless parameter~$\eta$. Since the scale $\Lambda = 0.72$ GeV together with the renormalized
quark masses $m_{u/d}(19 \,\mbox{GeV}) = 3.7$ MeV are fixed to experimental input (namely the pion decay constant $f_\pi$ and the pion mass),
only one free parameter $\eta$ remains to which many observables are insensitive within the range $1.6 < \eta < 2.0$. We discuss this point
further in Sec.~\ref{sec:results}.

In the following we will use Eq.~\eqref{couplingMT} both in the three-body and quark-diquark BSEs in order to systematically
compare the results of the two frameworks using the same underlying basis. In both cases one first solves  the DSE for the quark propagator,
thus determining the quark dressing functions in the complex momentum plane. In the three-body framework one then directly proceeds to the
three-body Faddeev equation, where the kernel $\mathbf{K}$ is a gluon exchange diagram dressed with the effective
interaction~\eqref{couplingMT}. In the quark-diquark framework one has to make a detour by first determining the masses, Bethe-Salpeter amplitudes and propagators of the
diquarks using their BSEs with the same two-body kernel $\mathbf{K}$. The resulting diquark amplitudes and propagators together with the quark
propagators then serve as input for the quark-diquark BSE for baryons.

In order to solve Bethe-Salpeter equations, they are treated as eigenvalue
problems. One uses the baryon mass as an external parameter and solves for the
eigenvalues of the BSE matrix. If, for a certain baryon mass, one of the
eigenvalues equals unity the corresponding eigenvector gives the Bethe-Salpeter amplitude.
In order to extract the spectrum of excited states in the three-body Faddeev
framework we use an implementation of the Arnoldi algorithm~\cite{arpack-ug}.
Like all Krylov subspace methods, the Arnoldi algorithm allows one to find only a
small number of eigenvalues of large matrices, selected by a certain criterion.
The computational cost increases proportionally to the number of eigenvalues
sought. Unfortunately, we found that numerical artifacts (such as the
discretisation of integrals) results in the appearance of spurious
complex conjugated pairs of eigenvalues which make the search for real eigenvalues
extremely costly.

For the quark-diquark approach it turns out that a QR decomposition is possible due to the drastically reduced structure of the
baryon wave functions. This enables one to store the kernel matrix as a whole and extract all eigenvalues at once, thereby giving access to higher excited states without
the large numerical effort needed for the Arnoldi algorithm in the three-body system. We are thus in a position to present a much more
complete spectrum for this case.

\section{Results}\label{sec:results}

\subsection{Rainbow-ladder}

Before we embark on our discussion we would first like to make clear what we can expect from the rainbow-ladder approximation.
To this end it is useful to recapitulate what has been found in the meson sector of QCD. In the following we summarise a discussion
presented in more detail in the review \cite{review}.

It has been argued within Coulomb-gauge QCD that rainbow-ladder is especially good in the heavy quark region where, in fact, it
becomes exact in the limit of very heavy quarks \cite{Watson:2012ht}.
It seems reasonable to expect a similar simplification in Landau gauge used in this work.
Comparing rainbow-ladder results in the heavy quark regime~\cite{Blank:2011ha,Hilger:2014nma,Fischer:2014cfa} with those
in the light quark sector \cite{Alkofer:2002bp,Krassnigg:2009zh,Fischer:2014xha} supports this notion. In the heavy quark
regime one finds an overall reasonable agreement of the rainbow-ladder results with the experimentally measured bound states
beyond the $D\bar{D}$ thresholds. The agreement is especially good for the pseudoscalar and vector mesons including their ground
states and radial excitations. In the quark model these are the `$s$ wave' states with vanishing orbital angular momentum.
Of the remaining states, the scalars and axialvectors show the largest discrepancies~\cite{Krassnigg:2009zh}.

In the light meson sector the spin-dependent
parts of the quark-antiquark interaction kernel become even more important and the deficiencies of the (vector-vector) rainbow-ladder
interaction become apparent. Whereas the light pseudoscalar (non-singlet) mesons, governed by dynamical chiral symmetry breaking,
are automatically reproduced in the symmetry-preserving rainbow-ladder scheme, also the vector mesons are in good agreement with
experiment~\cite{Maris:1999nt}. However, large deviations occur for scalar and axialvector states (the `$p$ waves' in the
quark model), which are only remedied in much more intricate truncations \cite{Chang:2009zb,Williams:2015cvx}. In the language
of the quark  potential models (strictly valid only in the heavy quark region) evidence suggests that rainbow-ladder calculations
roughly reproduce the size of the spin-spin contact part of the potential and the spin-orbit part, but materially overestimates
the binding in the tensor part of the spin-spin interaction.

For baryons this has interesting consequences. In the quark-model language also the ground-state octet and decuplet baryons -- in
our case, the $N(1/2^+)$ and $\Delta(3/2^+)$ -- are quark-model $s$~waves and therefore we may expect rainbow-ladder to provide a reliable
framework for the nucleon, the $\Delta$ and their excitations. Other spin-parity channels, however, may be significantly affected
by rainbow-ladder deficiencies and we expect masses that are too small, similar to the meson case.

Let us first discuss the $N(1/2^+)$ and $\Delta(3/2^+)$ channels. For the ground states, results in the three-body
framework~\cite{Eichmann:2009qa,Eichmann:2011vu,SanchisAlepuz:2011jn} and the quark-diquark approach~\cite{Eichmann:2007nn,Eichmann:2008ef,Nicmorus:2008vb}
have been found in very good agreement with experiment. Also the mass evolutions with varying current quark mass or, correspondingly,
varying pion mass have been discussed already and compare well with results from lattice QCD.
The new element in our present work is that we are now in a position to also add the respective excited states.
In the three-body framework we find the masses
\begin{equation}
m_{N^*}= 1.45(5) \,\mbox{GeV} \,, \quad m_{\Delta^*}= 1.49(6) \,\mbox{GeV}
\end{equation}
at the physical point.
The first is close to the experimental $N(1440)$, the Roper,
and the second agrees with the lower edge of the range of PDG values for the $\Delta(1600)$~\cite{Agashe:2014kda}.
The systematic errors correspond to the range $1.6 \le \eta \le 2.0$
for the width parameter in the effective interaction~\eqref{couplingMT}.
We verified that these states are indeed the first radial
excitations by inspection of their Faddeev amplitudes, which display a node when plotted over one of the relative momenta
between the three quarks.

Due to the tremendous amount of CPU time involved in
extracting excited states in the three-body framework we are only able to give the first radial excitation in these channels.
By contrast, in the quark-diquark approach the complexity of the Faddeev amplitudes is considerably smaller and
enables us to calculate the full spectrum below $\sim 2$~GeV.
Here we extend the setup in Refs.~\cite{Eichmann:2007nn,Eichmann:2009zx} by implementing not only scalar and axialvector diquarks
but also the pseudoscalar and vector diquarks with $I(J^P)=0(0^-)$ and $1(1^-)$, respectively, as
they turn out to be quantitatively important for several states~\cite{Eichmann:2016jqx}.
The resulting masses are
\begin{equation}\label{roper-qdq}
m_{N^*}= 1.50(9) \,\mbox{GeV} \,, \quad m_{\Delta^*}= 1.73(12) \,\mbox{GeV},
\end{equation}
where the errors refer to the same $\eta$ variation as described above.
Although the $\eta$ dependence for these two states is considerably larger than in the three-body case,
both masses are still compatible with the PDG range.

\begin{figure}[t]
        \begin{center}
        \includegraphics[width=1\columnwidth]{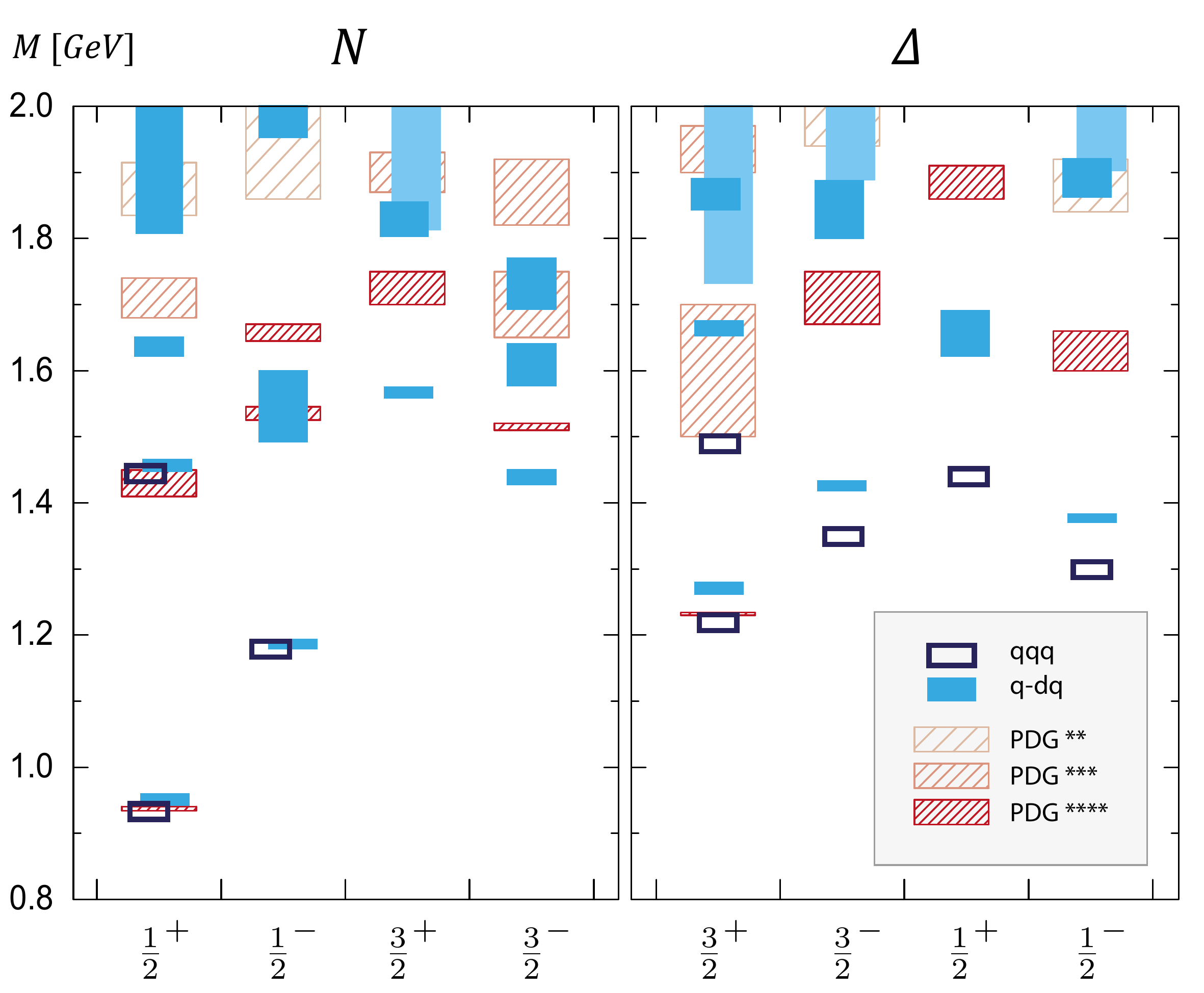}
        \caption{Nucleon and $\Delta$ baryon spectrum for $J^P=1/2^\pm$ and 
$3/2^\pm$ states determined within rainbow-ladder.
                 The three-body results (open boxes) are compared to the 
quark-diquark spectrum with full diquark content  for $\eta=1.7$ (filled boxes),
                 together with the PDG values and
                 their experimental uncertainties~\cite{Agashe:2014kda}.
                 The widths of our results represent an error estimate based on 
the corresponding eigenvalue curves
                (see appendix for details).} \label{spectrum}
        \end{center}
\end{figure}

As expected, in the remaining spin-parity channels we find masses that are 
significantly smaller than experiment.
In Fig.~\ref{spectrum} we compare the rainbow-ladder results to the PDG values.
The parity partner of the nucleon
is underestimated by $20\%$, leading to the wrong level ordering between
the Roper and the $N(1535)$. This is true both in the three-body and the 
quark-diquark approach;
note in particular that the three states (nucleon, Roper and parity partner) 
agree very well in the two frameworks.
In the nucleon channels with $J^P=3/2^\pm$ the situation is a little better but 
still not good; here we  only have
the quark-diquark results at our disposal because we were not able to extract 
corresponding states from the three-body equation.
A similar pattern can also be observed in the $\Delta$ sector, although the 
spread between the three-quark and quark-diquark results is somewhat larger.
For the ground and excited states with $J^P=3/2^+$ we  find again agreement with 
experiment, whereas the parity partners and the states with $J^P=1/2^\pm$ are
clearly off.

The situation is not improved by varying the only parameter in the system: 
within the range of $1.6 \le \eta \le 2.0$
we find the variations for the Roper and the excited $\Delta$ quoted above,
whereas all ground states but also the excited states in the other channels are 
less sensitive.
The nucleon and $\Delta$ ground states hardly move at all with $\eta$
and the typical variations for the remaining states are of the order of $\sim 
50$ MeV.

\begin{figure}[b]
        \begin{center}
        \includegraphics[width=1\columnwidth]{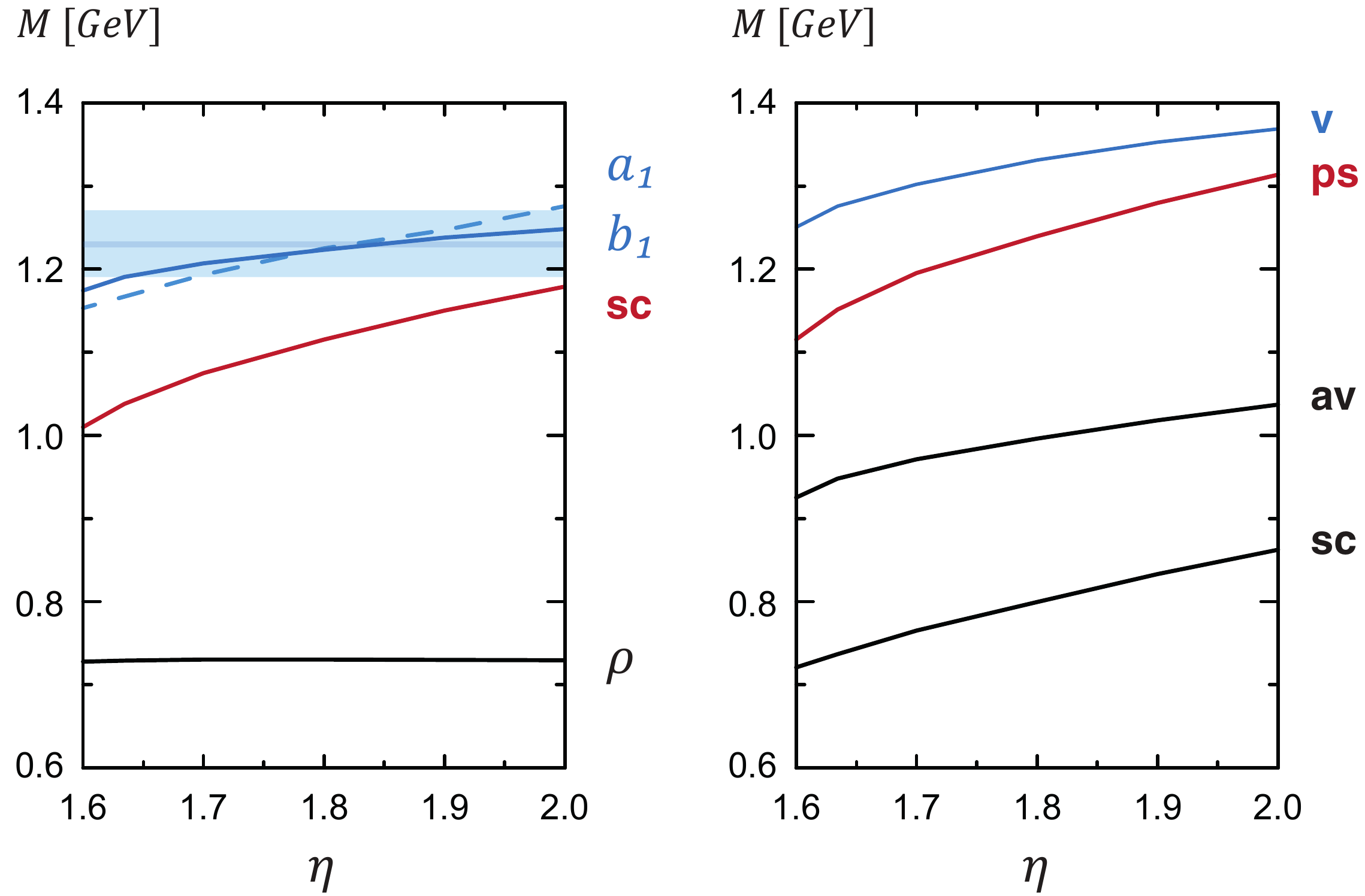}
        \caption{\textit{Left:} Vector, scalar and axialvector meson masses 
calculated in rainbow-ladder as functions of the $\eta$ parameter.
                 For the scalars and axialvectors we employed $c=0.35$ in order 
to shift the $a_1$ and $b_1$ masses towards
                 their experimental values (shown by the horizontal band).
                 \textit{Right:} Analogous plot for the diquark masses.} 
\label{meson-dq-masses}
        \end{center}
\end{figure}

There are a couple of important conclusions that can be drawn from these 
findings. First, both approaches essentially agree
with each other, especially in the nucleon channels and to a lesser extent also 
in the $\Delta$ channels,
which justifies the separable diquark approximation for the quark-quark
interactions inside baryons. The agreement with the experimental data in the 
`good' nucleon ($J^P=1/2^+$)
and $\Delta$ channels ($J^P=3/2^+$) furthermore
indicates that the omitted irreducible three-body forces indeed only play a 
minor role for the formation of baryon bound states
at least for these cases. Secondly, however, the disagreement in the `bad' 
channels (all the others)
between our results
and the experimental values confirms that parts of the interaction between the 
quarks are misrepresented
in the rainbow-ladder truncation underlying both frameworks. In the three-body 
approach this can be attributed
to the vector-vector character of the effective gluon exchange, which 
misrepresents some of the spin-dependent
parts of the interaction as discussed in the beginning of this section.

\subsection{A glimpse beyond rainbow-ladder}

To better understand the deficiencies of the rainbow-ladder approximation, it is
worthwhile to take a closer look at the underlying quark-diquark structure as it 
provides a link between the meson and baryon spectra.
After working out the Dirac, color and flavor structure the rainbow-ladder 
diquark BSEs~\eqref{bse-diquark}
become identical to their meson counterparts except for a factor 
$\nicefrac{1}{2}$ -- diquarks are `less bound' than mesons.
This entails that pseudoscalar, vector, scalar and axialvector mesons will 
exhibit similar features as their respective scalar, axialvector, pseudoscalar 
and vector diquark partners.
Pseudoscalar and vector meson properties are well reproduced in rainbow-ladder 
and thus the same can be said for scalar and axialvector diquarks
and the baryons made of them; hence these represent the `good' channels. On the 
other hand, the deficiencies of rainbow-ladder in the (`bad') scalar and 
axialvector meson channels
will translate into similar problems for pseudoscalar and vector diquarks.
Indeed, the typical mass scales obtained with rainbow-ladder calculations are 
about $800$~MeV for scalar diquarks and 1~GeV for axialvector diquarks,
whereas pseudoscalar and vector diquarks are only slightly heavier: about 1~GeV 
for pseudoscalar and 1.1~GeV for vector 
diquarks~\cite{Maris:2002yu,Eichmann:2016jqx}.
(The diquark masses also strongly depend on the $\eta$ parameter in contrast to 
the $\rho$ meson, cf.~Fig.~\ref{meson-dq-masses}.)
Hence, states with significant pseudoscalar and vector diquark content, such as 
the parity partners of the nucleon and the $\Delta$~\cite{Eichmann:2016jqx}, are 
expected to suffer from
too strong binding.

To remedy this problem, we follow the idea employed in 
Refs.~\cite{Roberts:2011cf,Chen:2012qr} in the context of the NJL-like 
contact-interaction model: we adjust the interaction strength in
the `bad' meson and diquark channels by a common constant factor $0 \le c \le 1$ 
that multiplies the interaction~\eqref{couplingMT}.
This increases the corresponding diquark masses and consequently
decreases their influence in the quark-quark interaction. Thus, dialling $c$ 
allows one to moderate the binding effects in the
quark-diquark BSE to the correct magnitude and thereby mimic beyond 
rainbow-ladder effects.
We gauge this factor in the meson sector by adjusting to the splitting of the 
vector/axialvector
parity partners, leaving the mass of the $\rho$ meson unchanged but increasing 
the masses of the $a_1$ and $b_1$; the corresponding value is $c \approx 0.35$  
(see Fig.~\ref{meson-dq-masses}).

This provides us with the following perspective: scalar and axialvector diquarks
contribute the underlying basis in \textit{all} baryon channels (for the 
$\Delta$ baryons only axialvector diquarks participate due to isospin 
combinatorics);
but except for the $N(1/2^+)$ and $\Delta(3/2^+)$ these baryons are additionally 
`contaminated' by pseudoscalar and vector diquarks
which are bound too strongly in rainbow-ladder.
We demonstrate this explicitly in the appendix
by considering a setup with all four diquarks included ($c=1$) and
one where we omit the pseudoscalar and vector diquarks ($c=0$). The `bad' 
diquarks  almost have no
impact on the `good' baryon channels whereas they substantially influence the 
masses in the other channels,
which leads to the unrealistically light baryon
masses in Fig.~\ref{spectrum}. This can also reconcile the contact-interaction 
studies of Refs.~\cite{Roberts:2011cf,Chen:2012qr},
where negative-parity baryon spectra were calculated using pseudoscalar and 
vector diquarks only,
with quantum-mechanical diquark 
models~\cite{DeSanctis:2011zz,Santopinto:2014opa} using scalar and axialvector 
diquarks only.
As we will see, one has to aim for a middle ground.

\begin{figure}[t]
        \begin{center}
        \includegraphics[width=1\columnwidth]{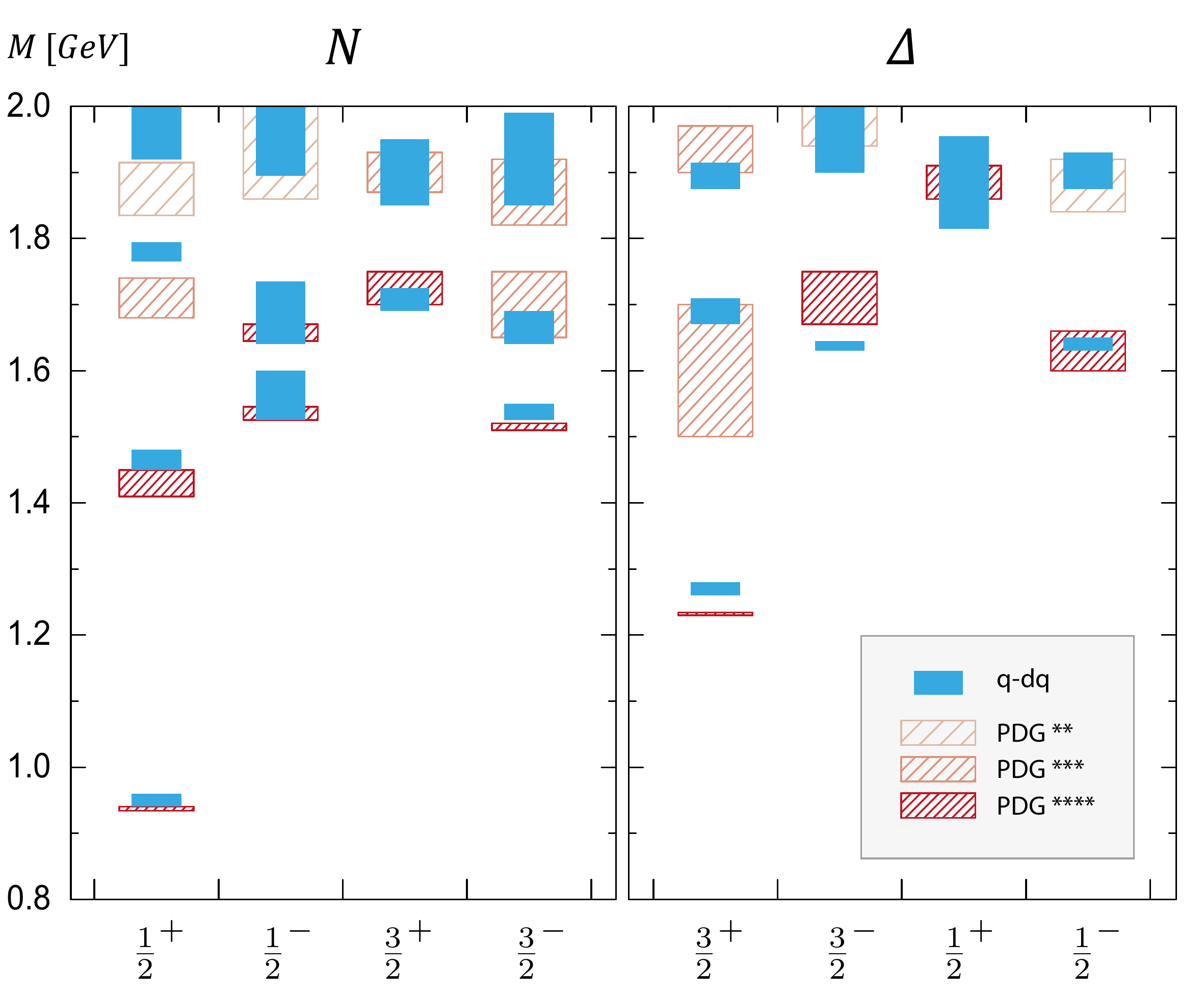}
        \caption{Nucleon and $\Delta$ spectrum with reduced strength in the
                pseudoscalar and vector diquark channels; see text for a 
detailed discussion. } \label{spectrum-2}
        \end{center}
\end{figure}

The resulting spectrum for $c=0.35$ is shown in Fig.~\ref{spectrum-2}. We find a 
drastic improvement
in the problematic channels, with hardly any changes in those that have been 
good before. The overall spectrum is
now in very good agreement with experiment, with a one-to-one correspondence of 
the number of observed to calculated
states in all cases and discrepancies below the $3\%$ level.
 Considering that there are only three relevant parameters involved, the scale 
$\Lambda$ fixed via $f_\pi$,
the factor $c$ fixed by the $\rho-a_1$ splitting, and the parameter $\eta$ with 
only a small influence on the spectrum, the
overall agreement is remarkable.

As an example, the level ordering between the Roper and the $N(1535)$ is now 
correct --
the latter had been polluted by the pseudoscalar and vector diquarks whereas the 
former was not.
Also the first radial excitation in the $N(3/2^+)$ channel nicely agrees with 
the experimental $N(1900)$,
which is a state that traditionally did not emerge from quark-diquark potential 
models~\cite{Nikonov:2007br}.

\begin{figure*}[t]
        \begin{center}
        \includegraphics[width=0.99\columnwidth]{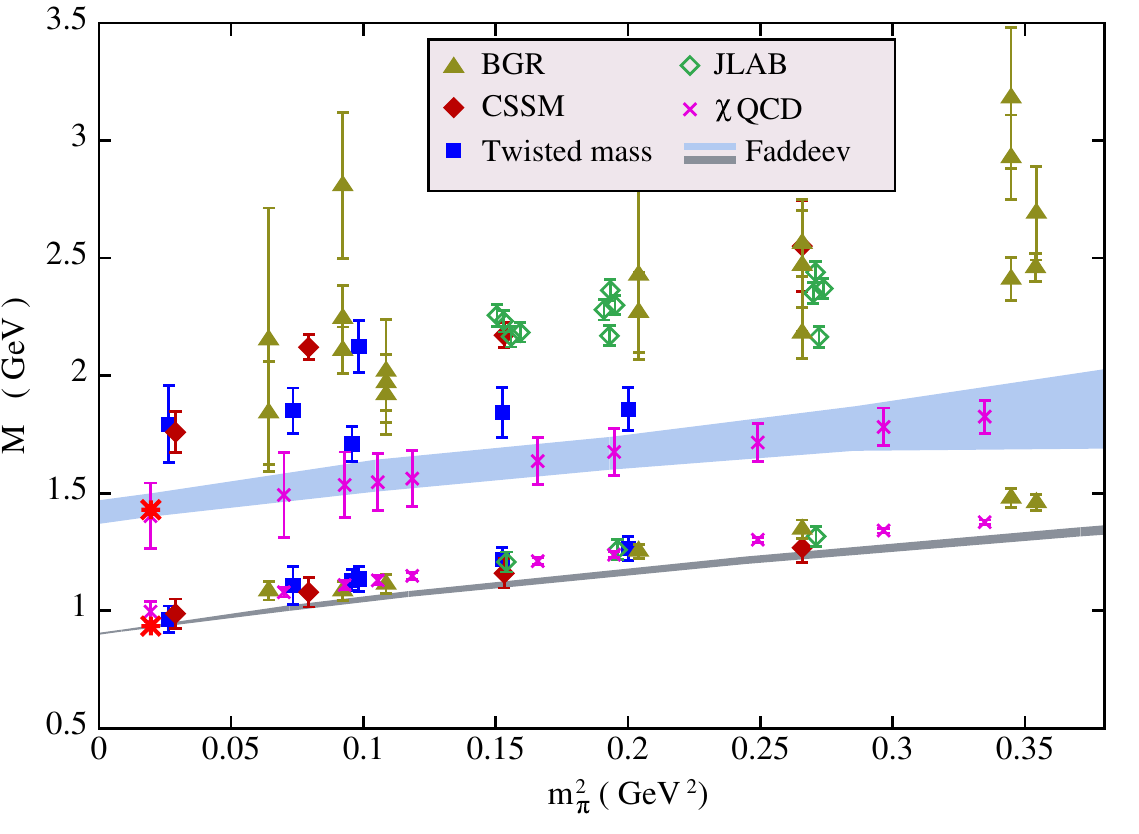}\hfill
        \includegraphics[width=0.48\textwidth]{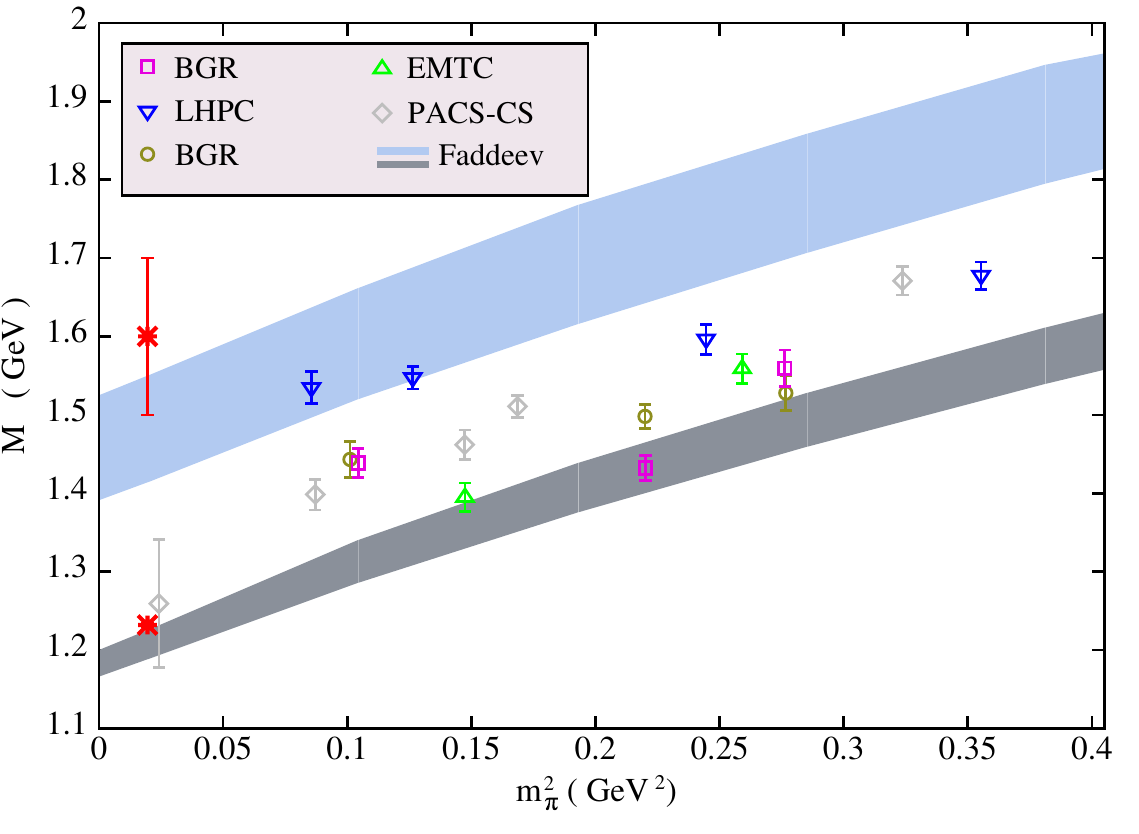}
        \caption{Mass evolution of the nucleon and Roper resonance (left) and 
the $\Delta$ and its first excited state (right).
        The bands are results obtained in the three-body rainbow-ladder Faddeev 
framework with $1.6 \le \eta \le 2.0$. In the left
        plot they are compared with lattice data for the nucleon and its first 
excitation~\cite{Edwards:2011jj,Mahbub:2010rm,Mahbub:2012ri,Alexandrou:2013fsu,
Alexandrou:2014mka,Alexandrou:2009hs,Liu:2014jua}. In the right plot, only 
lattice data for the $\Delta$ ground state are shown; data for the excited
$\Delta$ give values too high to be visible in the 
plot~\cite{Gattringer:2008vj,WalkerLoud:2008bp,Aoki:2008sm,Engel:2013ig}.} 
\label{nucleon_delta}
        \end{center}
\end{figure*}

Our findings are also interesting in view of the fact that the resulting baryons 
are still bound states without hadronic decay widths.
Ultimately the meson-baryon dynamics (which are beyond rainbow-ladder effects) 
will shift their poles into the complex plane and produce thresholds.
It is often assumed that this comes in combination with large attractive mass 
shifts or even a dynamical generation of resonances.
From our point of view two statements can be made in this respect:

(i) Beyond rainbow-ladder effects can compete nontrivially and they also affect 
$f_\pi$.
Because we fixed the scale~$\Lambda$ to reproduce the experimental pion decay 
constant,
those contributions that affect the masses and $f_\pi$ by the same amount would 
drop out from our plots and only the net effects remain visible, such as for 
example chiral non-analyticities.
To this end, elastic and transition form factors should provide much better 
signatures for `meson-cloud effects' because they are not affected by the scale 
setting.

(ii) Although rainbow-ladder generates bound states, one can still 
\textit{calculate} decay widths from their transition currents: for example,
 the $\Delta\to N\pi$ decay is the residue of the pseudoscalar $N\to\Delta$ 
transition form factor at the pion pole ($Q^2=-m_\pi^2$),
 and existing calculations yield quite reasonable values for such 
decays~\cite{Mader:2011zf,review}.
 Ultimately, these decay mechanisms would have to be backfed into the baryon  
bound-state equations
 and this is what would shift their T-matrix poles into the complex plane and 
thereby generate the desired widths.
 However, it would mainly represent a `correction' that comes on top of 
dynamically generating baryons as three-quark systems in the first place.
 Our results then suggest that the quark-gluon dynamics are indeed sufficient to 
produce all observed levels below 2 GeV,
whereas coupled-channel interactions would have comparatively mild effects that 
leave the real parts of the masses essentially unchanged
(or, alternatively, affect them all by a similar percentage together with 
$f_\pi$).

Unfortunately, in the three-body framework a simple change of the interaction 
that is selective between `good' and `bad'
contributions is not possible. Explicit diquark degrees of freedom no longer 
appear therein because the equation implicitly
sums over all diquarks.
Thus, in order to achieve similar results one would truly need to go beyond the
rainbow-ladder approximation.
First such calculations are available for ground-state nucleon and $\Delta$
masses in the three-body framework \cite{Sanchis-Alepuz:2015qra},
although the technical and numerical effort to solve the corresponding
Faddeev equations is substantial.
A convenient gauge of the quality of such truncations is of course already
the meson spectrum because scalar and axialvector mesons need
to acquire a larger mass. A systematic truncation based on the 3PI formalism 
that generates this effect has been
discussed recently in~\cite{Williams:2015cvx}. Making it available also in the 
three-body framework
is a major task that is left for the future.

In any case, our analysis shows that the $N(1/2^+)$ and $\Delta(3/2^+)$ as well 
as their first radial excitations
are insensitive to the addition or removal of pseudoscalar and vector diquarks. 
Our line of arguments
then suggests that they should also be stable when going beyond rainbow-ladder, 
and hence their rainbow-ladder results can be considered reliable.

\subsection{Current-mass evolution}

Returning to the three-body equation, we show in
Fig.~\ref{nucleon_delta} the evolution of the calculated ground and first 
excited states of the nucleon and $\Delta$ with the squared pion mass
and compare them with lattice QCD results. The shaded bands reflect the 
variation
of our results with the parameter $1.6 \le \eta \le 2.0$. Whereas the ground 
states are almost independent of $\eta$, the
excited states show a larger variation which is similar to observations made in 
the meson spectrum, see
e.g.~\cite{Krassnigg:2009zh,Fischer:2014cfa}.
Below a pion mass of about $m_\pi^2 = 0.15$ GeV$^2$ the masses rise 
approximately quadratically with $m_\pi$, whereas
above this value the behaviour eventually becomes linear. For the nucleon this 
is in agreement with chiral perturbation theory
in the region where the chiral expansion can  be safely applied, see e.g. 
\cite{Alvarez-Ruso:2013fza} and references therein,
and with the linear behaviour seen in lattice QCD for larger pion masses.

Whereas the lattice data on the mass evolution of the nucleon nicely agree with 
each other, the situation
for the Roper is somewhat different and the results are much more scattered. We 
find that, within error bars, our results
agree very well with those from the $\chi$QCD group~\cite{Liu:2014jua}.
On the other hand, it has been recently argued~\cite{Leinweber:2015kyz}
that the differences of the Roper mass evolution results from the JLab 
HSC~\cite{Edwards:2011jj}, CSSM~\cite{Mahbub:2010rm,Mahbub:2012ri}
and Cyprus groups~\cite{Alexandrou:2013fsu,Alexandrou:2014mka} visible in the 
plot can be accounted for and brought to consensus
with each other, and that this consensus deviates from the $\chi$QCD result. In 
general it seems fair to state that there may
not be an overall agreement in the lattice community concerning the status of 
the Roper and it will be very interesting to
see how this issue will be clarified in the future. In the $\Delta$ channel the 
situation is even less clear. While for the
ground state reasonable agreement may be claimed between our results and the 
lattice evolution, the existing lattice data for the
excited state gives values too high to be visible in our plot. This situation 
needs to be resolved.

\begin{table}[t]
\begin{tabular}{l @{\qquad}| @{\quad} c @{\quad} c@{\quad} c@{\quad} c}
\toprule
                   \%      & $N$      & $N^\ast(1440)$  & $\Delta$  & 
$\Delta^\ast(1600)$    \\\hline
\midrule
                $s$ wave     & $66$     &  $15$     & $56$    & $10$             
      \\
                $p$ wave     & $33$     &  $61$     & $40$    & $33$             
        \\
                $d$ wave     & $1$      &  $24$     & $3$     & $41$             
          \\
                $f$ wave     & $-$      &  $-$      & $<0.5$  & $16$             
          \\
\bottomrule
\end{tabular}
\caption{Magnitude of the orbital angular momentum contributions for the 
nucleon, Roper, $\Delta$, and excited $\Delta$.}\label{n-structure}
\end{table}

It is also interesting to study the internal structure of the radially excited 
states. As discussed above, the tensor
structures for the nucleon and the $\Delta$ can be grouped in $s$, $p$, $d$, and 
$f$ waves and their relative importance
can be assessed by their relative weight in the normalisation procedure of the 
Faddeev amplitudes. Our results are shown
in Table~\ref{n-structure}. Whereas the ground-state nucleon and $\Delta$ are 
dominated by $s$-wave components accompanied
by sizeable $p$-wave contributions, the excited states have a different internal 
structure. The Roper is dominated by
$p$-wave components and even the $d$ waves are stronger than the $s$-wave 
contribution. For the excited $\Delta$ baryon it
is even the $d$ waves that dominate and sizeable $f$-wave contributions are 
stronger than the $s$ waves. It will be very
interesting to probe these different internal structures in elastic and 
transition form factor calculations, which will
be the subject of future work.

\section{Conclusions}\label{sec:conclusions}

We have presented and discussed first results for the ground and excited states 
of the light baryon spectrum
in a Dyson-Schwinger/Bethe-Salpeter/Faddeev framework based on a 
momentum-dependent rainbow-ladder truncation.
Using the same underlying effective quark-gluon coupling, we systematically 
compared results for the three-body and the
quark-diquark framework. Due to restrictions in terms of numerical complexity a 
full spectrum of excited
states in all channels could only be obtained in the latter, whereas in the 
three-body framework we were
restricted to the Roper and the $\Delta(1600)$.

Our results can be summarised as follows. First, for those ground and excited
states that are numerically accessible in both frameworks we found reasonable to 
good agreement between the approaches.
Second, the quark-diquark spectrum for $J^P=1/2^\pm$ and $3/2^\pm$ states of 
nucleon and $\Delta$ type agrees with the one from
the PDG~\cite{Agashe:2014kda} on a qualitative basis; once well-understood 
deficiencies in the rainbow-ladder
framework are remedied, the resulting spectrum is even in very good quantitative 
agreement with experiment.
In particular, we could reproduce the masses of all experimental
states below 2~GeV at the $3\%$ level, including the correct level ordering 
between the Roper and the parity partner of the nucleon.
This agreement is highly non-trivial and relies on intricate
and channel-dependent cancellations between the effects of different diquarks, 
which are nevertheless fully controlled by one
parameter only. It thus appears that the quark-diquark picture of baryons (with 
fully momentum-dependent diquarks) works
very efficiently at least for the states considered in this work.
It will be interesting to see whether a similar agreement is possible for the 
decay widths; a first study
of $g_{N \Delta \pi}$ discussed in Ref.~\cite{Mader:2011zf} indeed points in 
such a direction.

In general, we did not find any arguments from the spectrum calculated so far 
that could distinguish between a three-body
or a realistic, momentum-dependent quark-diquark picture of baryons. It remains 
to be seen whether this is still
the case in an extended calculation beyond the channels presented here and for 
higher excitations. Note, however, that subleading components of
electromagnetic form factors  may indeed be able to discriminate between the 
two~\cite{review}.

A special case of further interest is still the Roper. Within our framework we 
find that it is well represented as the first
radial excitation of the nucleon with a mass close to experiment. This may 
indicate that potential
quantitative corrections stemming from beyond rainbow-ladder contributions can 
either be absorbed in the scale setting,
or that they are small and only marginally affect the real part of the mass.
In order to shed light on this question we also determined the mass evolution of 
the Roper with varying current-quark mass or,
correspondingly, varying pion mass up to the region where dynamical 
coupled-channel effects should no longer play a role. Unfortunately,
the comparison of the mass evolution with lattice data remains inconclusive due 
to the spread in
the available lattice results from different groups. The good agreement with the 
results of $\chi$QCD may (or may not) be
accidental and this issue needs to be explored further in the future.

\vspace*{0.2cm}
\noindent {\bf Acknowledgements}
We thank R.~Alkofer and R.~Williams for a critical reading of the manuscript.
This work has been supported by an Erwin Schr\"odinger fellowship J3392-N20 from 
the Austrian Science Fund, FWF,
by the Helmholtz International Center for FAIR within the LOEWE program of the 
State of Hesse,
by the DFG collaborative research center TR 16 and the BMBF project 05H15RGKBA.

\begin{appendix}

\section{Eigenvalue spectra}

In the following we provide details on the eigenvalue spectra extracted from the 
quark-diquark calculation.
Bethe-Salpeter equations such as the quark-diquark BSE in Eq.~\eqref{qdq-bse} 
are homogeneous eigenvalue equations:
$\mathbf{K\,G_0}\,\mathbf\Gamma = \lambda\,\mathbf\Gamma$, where $\mathbf{G_0}$ 
abbreviates the combination of quark and diquark propagators
and we introduced an artificial eigenvalue $\lambda(P^2)$. Since $P^2=-M^2$ is 
an external parameter,
the eigenvalue spectrum $\lambda_i(P^2)$ of the kernel $\mathbf{K\,G_0}$ allows 
one to
read off the masses $M_i$ of the ground and excited states from the 
intersections $\lambda_i(P^2=-M_i^2) = 1$.

\begin{figure*}[t]
        \begin{center}
        \includegraphics[width=0.98\textwidth]{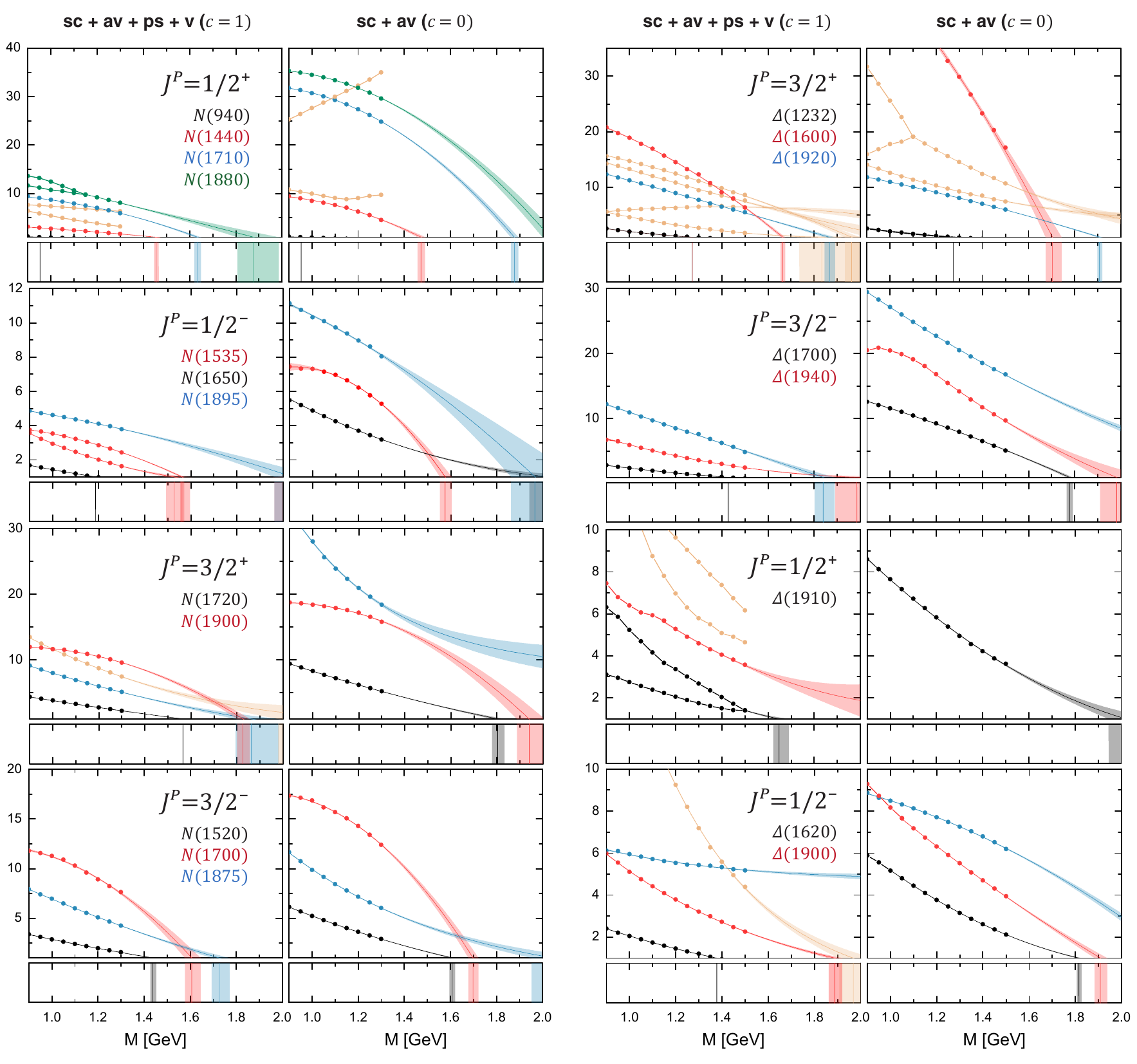}
        \caption{Inverse eigenvalue spectra $1/\lambda_i(M)$ for the four nucleon (\textit{left}) and four $\Delta$ channels (\textit{right}).
                 In both cases the left panels show the results with all diquarks included and the right panels those with scalar and axialvector diquarks only.} \label{ev-spectrum}
        \end{center}
\end{figure*}

In Fig.~\ref{ev-spectrum} we plot the resulting (inverse) eigenvalue spectra $1/\lambda_i$ as functions of $M$ for the eight baryon channels we investigated.
The parameter $c$ controls the interaction strength in the pseudoscalar and vector diquark channels.
The left columns show the results with all diquarks included ($c=1$) and the right columns with scalar
and axialvector diquarks only ($c=0$). The spectra for $c=0.35$ cannot be calculated directly because the first complex conjugate pole pair
in the quark propagator defines a parabola mass limit $m_P \sim 0.55$ GeV,
where meson and diquark masses above $2m_P$ can only be extrapolated (unless we performed residue calculus); cf. the discussion around Figs. 3.8--3.9 in Ref.~\cite{review}.
This happens for $c \lesssim 0.75$; below that value the pseudoscalar and vector diquark masses exceed the
contour limit $2m_P$. We find, however, that the masses are approximately linear in $c$ which allows us to perform
linear interpolations for the baryon masses between the cases $c=1$ and $c=0$.

Although the QR algorithm returns all eigenvalues, only the first ten or so are stable within our numerical accuracy whereas the remaining ones require increasing resolution.
The quark-diquark equation defines another contour limit, namely the sum of the quark parabola mass $m_P$ and the lowest diquark mass in the system:
$m_P+m_\text{sc} \approx 1.35$ GeV for the nucleons with isospin $I=\nicefrac{1}{2}$ and $m_P+m_\text{av}\approx 1.55$ GeV for the $\Delta$ baryons with $I=\nicefrac{3}{2}$.
Baryon masses above those limits are extrapolated as shown in Fig.~\ref{ev-spectrum}.
The dots are the calculated eigenvalue spectra and for their extrapolation to $\lambda_i=1$ we used polynomial fits with $80\%$ confidence bands.

The results in Fig.~\ref{ev-spectrum} lead us to the following observations.
The case $c=1$ with all diquarks included generally produces rather dense eigenvalue spectra, leading to the masses in Fig.~\ref{spectrum}
which are typically too low compared to experiment. By comparison,
the eigenvalues for $c=0$ are rather sparse and produce states that are too high.
From the directly calculable cases between $c=1 \,\dots \,0.75$ we find that the eigenvalue curves gradually expand when lowering $c$, resulting in just a few eigenvalues at $c=0$
that are relevant for states below $2$ GeV. The eigenvalues shown for $c=0$ typically exhaust the depicted plot range whereas for $c=1$ there would be
further higher-lying curves which we do not show because they extrapolate to masses above 2 GeV.

In all channels we obtain both real and complex conjugate eigenvalues, although the imaginary parts are small
and shrink with the numerical accuracy so their complex nature is presumably just a numerical artifact. What occasionally happens, however, is that complex conjugate eigenvalues
can branch into two real ones, as is visible in the $N(\tfrac{1}{2}^\pm)$, $\Delta(\tfrac{3}{2}^+)$ and $\Delta(\tfrac{1}{2}^+)$ channels.
This does not appear to change with better numerics but it usually also
does not affect the mass extraction, with the exception of the $N(1535)$ where two such branches extrapolate to a common point,
and the $N(1880)$ and $\Delta(1910)$ where we averaged over the branches to perform the extrapolation.

The parity partner of the nucleon is an interesting case also for another reason: for $c=1$ the largest eigenvalue (or smallest inverse eigenvalue) produces a state at $\sim 1.2$~GeV, whereas the same
eigenvalue for $c=0$ extrapolates to 2 GeV. The intercept at $c=0.35$ generates two nearby states which are also seen experimentally; our analysis suggests
that it is actually the second state at $c=1$ that should be identified with the $N(1535)$. In all other cases
the sensitivity to the pseudoscalar and vector diquarks is less severe.
Observe in particular that the nucleon and $\Delta$ themselves, together with their first excitations
including the Roper resonance, are almost insensitive to the pseudoscalar and vector diquark content.

The results shown here (and the corresponding Figs.~\ref{spectrum} and~\ref{spectrum-2}) correspond
to the value $\eta=1.7$ in the effective interaction~\eqref{couplingMT}.
We repeated the analysis for different values up to $\eta=2.0$ but the eigenvalue curves do not materially change
(although the diquark masses vary considerably in this range, cf.~Fig.~\ref{meson-dq-masses})
and
the resulting baryon spectra remain similar. Notable exceptions are the nucleon and $\Delta$ excitations in the first row of Fig.~\ref{ev-spectrum},
where the Roper resonance and $\Delta(1600)$ move within the range in Eq.~\eqref{roper-qdq}.
Furthermore, those eigenvalues that split into two branches are found in all cases although the branching may set in at different values for $M$
(or even in the reverse direction), which also helps us identify and connect them between the limits $c=1$ and $c=0$.

Finally, we also calculated the eigenvalue spectra for larger pion masses but also here we did not find any qualitative changes:
the states evolve similarly to those in Fig.~\ref{nucleon_delta} with the current-quark mass,
and the level ordering between the `Roper' and `$N(1535)$' remains intact.

\end{appendix}

\bibliography{roper2}

\end{document}